# Real-time experiment-theory closed-loop interaction for autonomous materials science

Haotong Liang[1], Chuangye Wang[1], Heshan Yu[1,*], Dylan Kirsch[1], Rohit Pant[1], Austin McDannald[2,†], A. Gilad Kusne[1,2,#], Ji-Cheng Zhao[1], Ichiro Takeuchi[1,3]

[1] Department of Materials Science and Engineering, University of Maryland, College Park, MD 20742, USA
[2] National Institute of Standards and Technology, Gaithersburg, MD 20889, USA
[3] Maryland Quantum Materials Center, University of Maryland, College Park, MD 20742, USA
*Present address: School of Microelectronics, Tianjin University, Tianjin, China 300072
† ORCID: 0000-0002-3767-926X
# ORCID: 0000-0001-8904-2087

## Abstract

Iterative cycles of theoretical prediction and experimental validation are the cornerstone of the modern scientific method. However, the proverbial "closing of the loop" in experiment-theory cycles in practice are usually *ad hoc*, often inherently difficult, or impractical to repeat on a systematic basis, beset by the scale or the time constraint of computation or the phenomena under study. Here, we demonstrate Autonomous MAterials Search Engine (AMASE), where we enlist robot science to perform self-driving continuous cyclical interaction of experiments and computational predictions for materials exploration. In particular, we have applied the AMASE formalism to the rapid mapping of a temperature–composition phase diagram, a fundamental task for the search and discovery of new materials. Thermal processing and experimental determination of compositional phase boundaries in thin films are autonomously interspersed with real-time updating of the phase diagram prediction through the minimization of Gibbs free energies. AMASE was able to accurately determine the eutectic phase diagram of the Sn-Bi binary thin-film system on the fly from a self-guided campaign covering just a small fraction of the entire composition – temperature phase space, translating to a 6-fold reduction in the number of necessary experiments. This study demonstrates for the first time the possibility of real-time, autonomous, and iterative interactions of experiments and theory carried out without any human intervention.

Autonomous experimentation driven by active learning —the field of machine learning (ML) dedicated to dynamically steering the sequence of experiments —is poised to revolutionize scientific discovery at large(*1–11*). At the core of active learning is Bayesian optimization (BO), whose implementation is increasingly popular in research laboratory settings for rapid navigation of hitherto unknown response functions of physical systems under study. BO is a sequential decision-making process grounded in the framework of Bayesian inference that iteratively updates the underlying with the observed evidence at each iteration(*12*). Further integration of automated experiment platforms with BO based active learning algorithm has led to the dawn of fully autonomous research systems(*6*, *13–16*), where dynamic decision-making vis-a-vis the task to be performed at each cycle is now squarely in the hands of robot scientists(*17*).

One of the most challenging yet worthy areas of autonomous research systems is materials discovery. Despite technical difficulties associated with live integration of synthesis and characterization tasks, the inherently iterative nature of materials exploration cycles (as exemplified in the Edisonian approach) is highly conducive to closed-loop autonomous systems, and they have been successfully demonstrated for synthesis of inorganic quantum dots(*18*) and optimization of organic solar cells(*19*). As a major benefit of autonomous systems, there is a significant reduction in the overall number of necessary experimental iterations to arrive at the optimum solutions in live runs. An essential task associated with materials discovery is the rapid mapping of compositional phase diagrams and identifications of phase boundaries. High-throughput mapping of phase diagrams has successfully led to discoveries of various functional materials(*20*, *21*), and a Bayesian autonomous workflow for streamlined mapping of phase boundaries in combinatorial libraries has previously led to the discovery of a best-in-class phase change memory material(*22*).

From electronic structure calculations and molecular dynamic simulations to predictions of structural materials and quantum phenomena, the field of computational materials science has witnessed staggering advances in recent years, which in turn have led to the era of high-throughput computational materials design and screening. Despite such advances, there naturally still remains a divide between theory and experiment. To scale the divide, there have been efforts to incorporate computational models into the autonomous workflow by including steps where separately-prepared or existing computational data are compared with experiment in order to guide the next iteration (*1*, *23*, *24*). In the framework of Bayesian optimization, the incorporation of physical models to the probabilistic model, or specifically, Gaussian process (GP), is typically implemented at the level of kernel functions(*25*), mean function(*26*), inputs and outputs(*24*), and acquisition function(*27*). Dynamic modification of the physical prior or theory as an iterative part of the autonomous closed-loop is still largely limited in scope. Exciting yet challenging are live autonomous cycles involving materials synthesis where there is time-constraint due to kinetics and, as a result, there is experimental urgency associated with chemical reactions taking place in real time.

We hereby report the successful demonstration of the Autonomous MAterials Search Engine (AMASE) where alternating live interaction of experimental observation and computational modeling is driven by Bayesian active learning informing and correcting each other without interruptions on the fly. In particular, we have implemented the system for the task of self-charting a thin-film eutectic phase diagram. The AMASE system performs threaded cyclical tasks of composition selection via active learning, x-ray diffraction measurement and its analysis, thermal processing of samples, interspersed with GP classification for phase boundary determination and thermodynamic calculations. The continuously updating live computation of Gibbs free energies combined with Bayesian autonomous experiments has allowed us to arrive at the complete thin-film phase diagram after sampling just a small fraction of the entire composition-temperature phase space. One live run constructing the Sn-Bi thin-film phase diagram is completed in just over 8 hours in an ambient lab environment, translating to **6-fold reduction in the total number of experiments compared** to an exhaustive and time-consuming grid mapping (in the same range), which is undesirable in air for materials containing such common yet volatile chemical elements.

Thermodynamics is a pervasive science fundamental to virtually all scientific disciplines. CALPHAD (CALculation of PHAse Diagram as implemented in the Thermo-Calc (TC) software(*28*)) is a naturally iterative method of adjusting the parameters that describe the Gibbs energies of various phases in a multi-element materials system to construct a thermodynamic phase diagram model as a best fit to the available experimental phase equilibrium data. We perform continuous live updating of the computed phase diagram (**Fig. 1a**), which in turn directly informs the next composition and temperature to experimentally determine the locations of phase boundaries (i.e., solvus and liquidus). The experiment is x-ray diffraction (XRD) determinations of a thin-film composition spread, which encompasses a segment of the Sn-Bi binary compositional phase diagram (**Fig. 1**), undergoing variable temperature conditions (see Supplementary Section 7 for experimental setup). More details of CALPHAD models are given in Section 4 of the Supplementary Materials.

While the bulk Sn-Bi phase diagram is well-documented(*29–32*), it is well known that thin films in general can display significant departure in thermodynamic behavior from their bulk counterparts. Thin films possess inherent two-dimensional structures placing stringent limitations on grain size due to diffusion kinetics and grain growth, accompanied by high rates of evaporation as well as stress from the substrate, which can also shift phase diagrams. The specific scenario for the task at hand is to decipher the Sn-Bi thin-film (≈0.5 μm thick) phase diagram, so that it can serve as a processing blueprint for fabrication of thin-film devices incorporating superconducting $Sn_xBi_{1-x}$ (*33*).

Structural phase mapping using composition spreads has become a lynchpin for exploration of advanced materials(*22*, *34*, *35*) and for probing mechanisms of exotic physical phenomena(*36*), as phase diagram navigation is central to the process of search and discovery. The marriage of the high-throughput platform with active learning then is a natural one, as an acquisition function can serve as an efficient beacon for steering across a well-ordered array of samples, eliminating the necessity to exhaustively comb through the entire composition map.

In order to showcase the ability of the approach to accurately create the entire phase diagram from just a limited sampling of experimental data and computation, we have deliberately prepared a thin-film composition spread which covers only a small fraction of the Sn-rich side of the Sn-Bi phase diagram, namely $0.71 < x < 0.95$ in $Sn_xBi_{1-x}$ (**Fig. 1b**). As general features of any binary phase diagram, the only assumption we make prior to the run is that the solvus, solidus, and liquidus lines lie "somewhere" within this region of the phase diagram, consistent with the bulk phase diagram, and it is the task of AMASE to accurately locate them in the present thin-film system. **Fig. 1c** is a photo of the experimental setup, where the composition spread wafer is placed on a variable-temperature diffractometer with position control, so that different composition spots and temperatures can be dialed in by AMASE via remote control from a main computer. Because of the irreversible nature of the thermal process here, we start the mapping at a low/ambient temperature and continue to move up in temperature as dictated by active learning.

Previously, we have implemented a number of different analysis techniques to automatically detect and quantify peak features from diffraction patterns while removing background noise(*37*, *38*). Here, we have developed a materials-agnostic ML analysis protocol that can identify diffraction peaks and monitor peak shifting in a series of diffraction patterns as a function of various parameters. Based on the You Only Look Once (YOLO) object detection model from computer vision(*39*, *40*), we developed a 1D YOLO model that can quickly capture 2θ positions and full widths at half maximum (FWHM) of all major diffraction peaks from an XRD pattern and track their change in intensity and position as the pattern evolves as a function of composition (due to changing chemical pressure) and temperature (due to thermal expansion). (see Supplementary Sections 1, 2, and 3 for more details on the modified YOLO model). The key task of the model is to help identify the phase boundary, where the intensity of a main tracked peak decreases to below the background noise level.

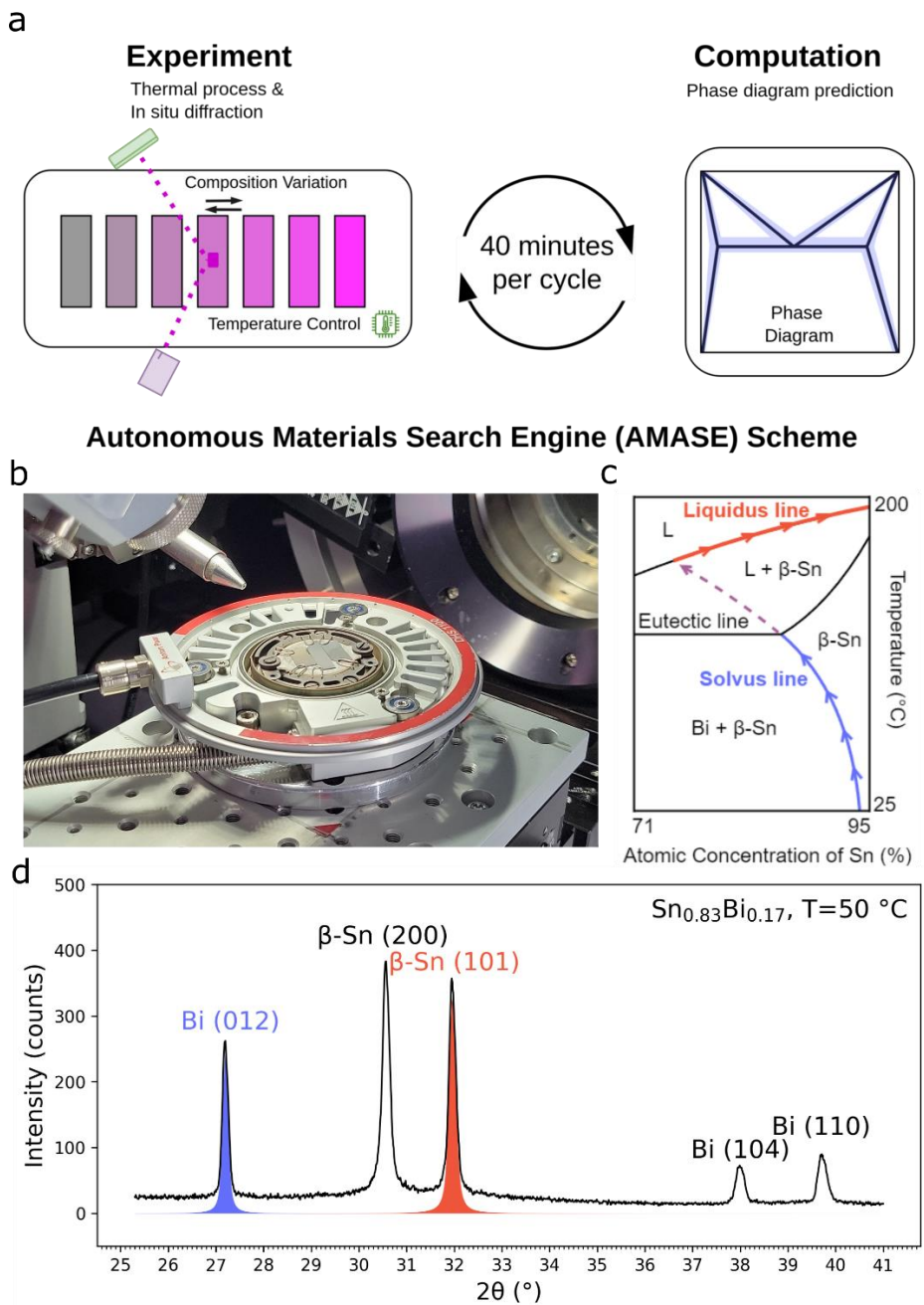

Fig. 1. AMASE (autonomous materials search engine) is applied to the mapping of the Sn-Bi thin-film phase diagram. a, The AMASE performs real-time experiment-calculation interaction for phase diagram determination. b, A photo of the experimental setup: a thin-film composition spread is mounted on a variable temperature stage inside an in-house x-ray diffractometer (Bruker D8) equipped with a scanning stage. c, The region and general features (solvus and liquidus lines indicated by blue and orange lines, respectively) of the thin-film eutectic binary (Sn-Bi) phase diagram AMASE is tasked to find and track experimentally on a composition spread focused on the Sn-rich region

(atomic concentration of Sn of 71 % to 95 %). The eutectic point composition to be determined by the AMASE process is assumed to be outside the composition range of the spread. d, XRD diffraction pattern of $Sn_{0.71}Bi_{0.29}$ at T = 50 °C. Key diffraction peaks used for tracking are indicated.

We initialize the AMASE process by measuring and analyzing the XRD of endpoint compositions of the spread, namely at $Sn_{0.71}Bi_{0.29}$ and $Sn_{0.95}Bi_{0.05}$ ($Sn_xBi_{1-x}$) at room temperature (**Fig. 1d**). The peaks identified with the modified YOLO model are indexed via the search/match algorithm using Inorganic Crystal Structure Database (ICSD)(*41*). As the most prominent characteristic peaks, the Bi (012) peak and the β-Sn (101) peak are chosen here for finding and tracking the solvus line compositions and the liquidus line compositions, respectively.

Following the initialization, the AMASE workflow commences by launching an isothermal phase-boundary-search-routine for solvus composition identification at 50 °C as the starting low temperature, which was chosen, so that we cover both sides of the solvus line composition (between the β-Sn + Bi two-phase region and the single phase β-Sn region) (**Fig. 1b**) in the composition range of the spread at the starting point of the workflow.

At the core of the phase-boundary-search-routine is the variational Gaussian process classifier (VGPC), which is a robust non-parametric probabilistic ML model(*42*, *43*) that uses CALPHAD evaluated phase boundary as its prior to predict the true phase boundary in a probabilistic approach. At a fixed temperature, the inputs are the compositions and the corresponding intensity of the Bi (012) peak from the YOLO fit of diffraction patterns. The output is the probabilistic classification of the constituent phases at a given composition and temperature, which triggers the next XRD measurement at the composition location where the VGPC mean curve shows a drastic change (i.e., first-order phase transition), indicating the predicted position of the boundary. This cycle continues until convergence is reached, where at the predicted solvus composition, the Bi (012) peak (and thus the phase) disappears. In order to confirm convergence and signal the end of the routine at a fixed temperature, XRD measurements at both sides of convergent composition with $\Delta x = \pm 0.01$ are carried out.

**Fig. 2** shows this converging process of the phase-boundary-search-routine through multiple iterations at 130 °C. The red squares indicate locations on the spread where successive measurements have identified compositions where the Bi (012) peak is present, and thus are in the two-phase region. The sharp transition of the VGPC mean-curve indicates the phase boundary is at an atomic Sn concentration of 77 % (x = 0.77) at this temperature. On the left side of the boundary (i.e., lower x for $Sn_xBi_{1-x}$), x-ray measurements have confirmed absence of the Bi (012) peak indicated by purple squares. On average, the phase-boundary-search-routine takes six XRD iterations and 40 minutes to converge at a given temperature.

**Fig. 3** shows the autonomous interactive block diagram of the AMASE workflow. Once the solvus composition is experimentally identified at a particular temperature, this info is fed to CALPHAD which proceeds to calculate the entire phase diagram by minimizing the Gibbs free energy functions of three phases in the system, namely, β-Sn, Bi, and liquid here (See Supplementary Sections 4 and 6). Using the Redlich-Kister polynomial up to the second order as the two primary fitting parameters per phase in the Gibbs functions, the Thermo-Calc optimization is performed ten times to arrive at a phase diagram with uncertainty as to the position of the phase boundaries. This computational step typically takes less than one minute.

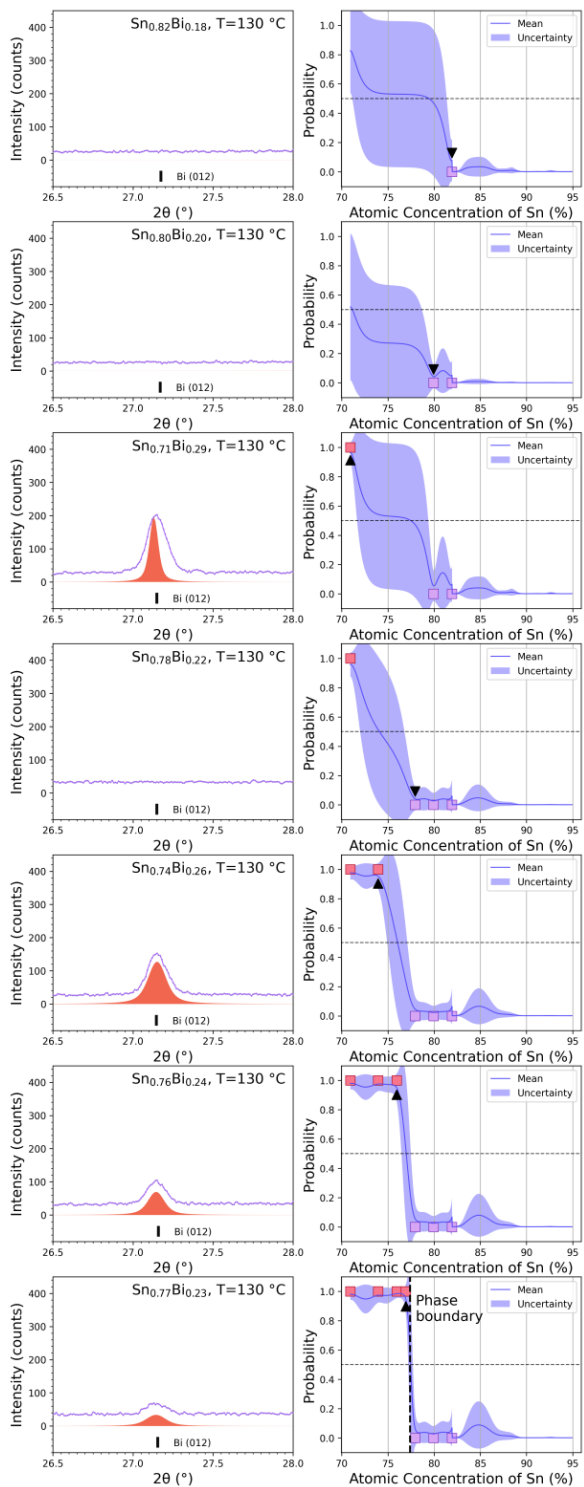

Fig. 2. Isothermal phase boundary (solvus) search process performed autonomously using the YOLO-based diffraction peak detection method (left) and accompanying variational Gaussian process classifier (VGPC) (right) across the $Sn_xBi_{1-x}$ composition spread (0.71 < x < 0.95). Here at 130 °C, the Bi (012) diffraction peak is tracked as composition is varied. The left panels show the diffraction peak with the YOLO fit after each measurement, and the right panels show the VGPC result after each measurement, where the updated boundary prediction with the uncertainty envelope (95 percent confidence intervals) is plotted. Red and purple squares indicate the XRD measurement result after each successive iteration plotted against composition denoting "left" and "right" side of the boundary, respectively. Going from the top panels to the bottom, this particular process took 5 measurements after the initial setup (top panels) to find the boundary, where a sharp boundary with minimal uncertainty was found at $Sn_{0.77}Bi_{0.23}$ (at 130 °C) as shown on the bottom right panel, and the diffraction pattern shows a peak below the noise level for $Sn_{0.78}Bi_{0.22}$ (left second from bottom).

As the key following step, AMASE then decides the next temperature to perform the phase-boundary-search-loop. The uncertainty in the calculated solvus line composition at different temperatures is used in the acquisition function in the exploration mode within the window (set to be 30 °C here) above the last temperature where the phase-boundary-search-routine was carried out. The size of the window is chosen such that at least five temperature levels would be surveyed through the entire range of temperature from 50 to 200 °C. This important step works in such a way as to eliminate the possibility of having to perform the experiment at every 10 °C increment, thus substantially reducing the overall number of isothermal boundary-search-routines. The temperature on the stage is raised to the next selected value, and the workflow continues with the VGPC boundary-search-routine at the composition given by the calculated phase diagram as the prior (see Supplementary Section 4.2), thus completing the experiment-computation thermodynamic phase modeling loop.

In this manner, the AMASE cycle walks up the solvus line driven by the experiment-calculation interactive loop until both VGPC and the CALPHAD models predict vanishing of the solvus phase boundary at a temperature within the 30 °C window above the sampled temperature. This signals the end point of the solvus line. Another Gaussian process classifier is implemented as the end-point detection scheme (see Supplementary Section 5), and upon confirmation of arrival at the eutectic temperature, the mapping of the solvus line is terminated.

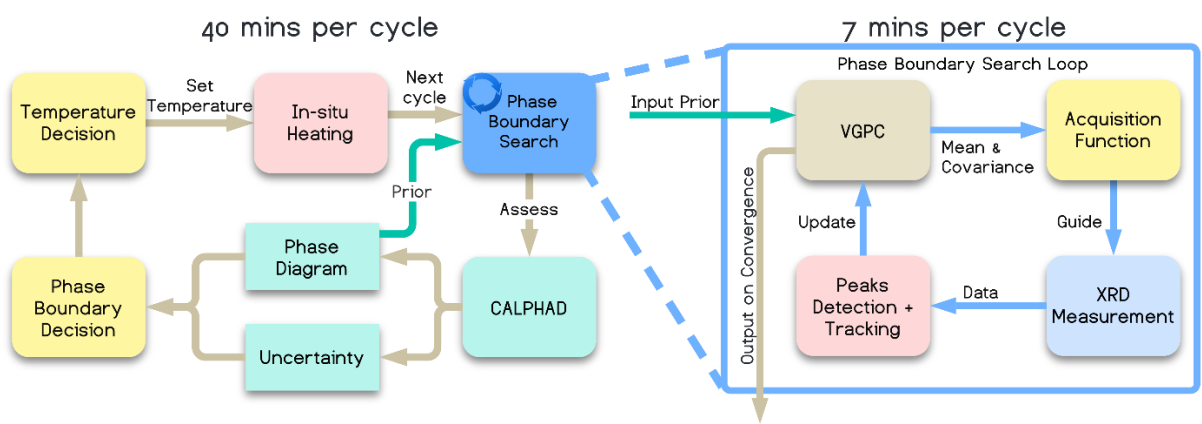

Fig. 3. The block diagram shows the steps in the closed-loop cycles starting at room temperature. The process at each temperature takes approximately 40 min. for performing x-ray diffraction at automatically selected few spots on a thin-film composition spread and GP classification to identify the solvus composition. The next temperature to sample is determined autonomously via active learning. After the eutectic isotherm is reached, the process continues

by jumping to find and track the liquidus line. For the present run, the entire process starts with a Phase Boundary Search Loop at 50 °C, with no initial input from CALPHAD or the bulk phase diagram to begin with.

As captured in the present AMASE block diagram (**Fig. 3**), the next step is to jump to the liquidus search-routine mode. The lowest temperature of the liquidus within the composition, as dictated by the most-recently updated thermodynamic database prior, becomes the starting temperature of the next phase-boundary-search-routine, which would undergo the same process as above. The process continues until 200 °C, which is the high temperature operation limit of the heater stage used.

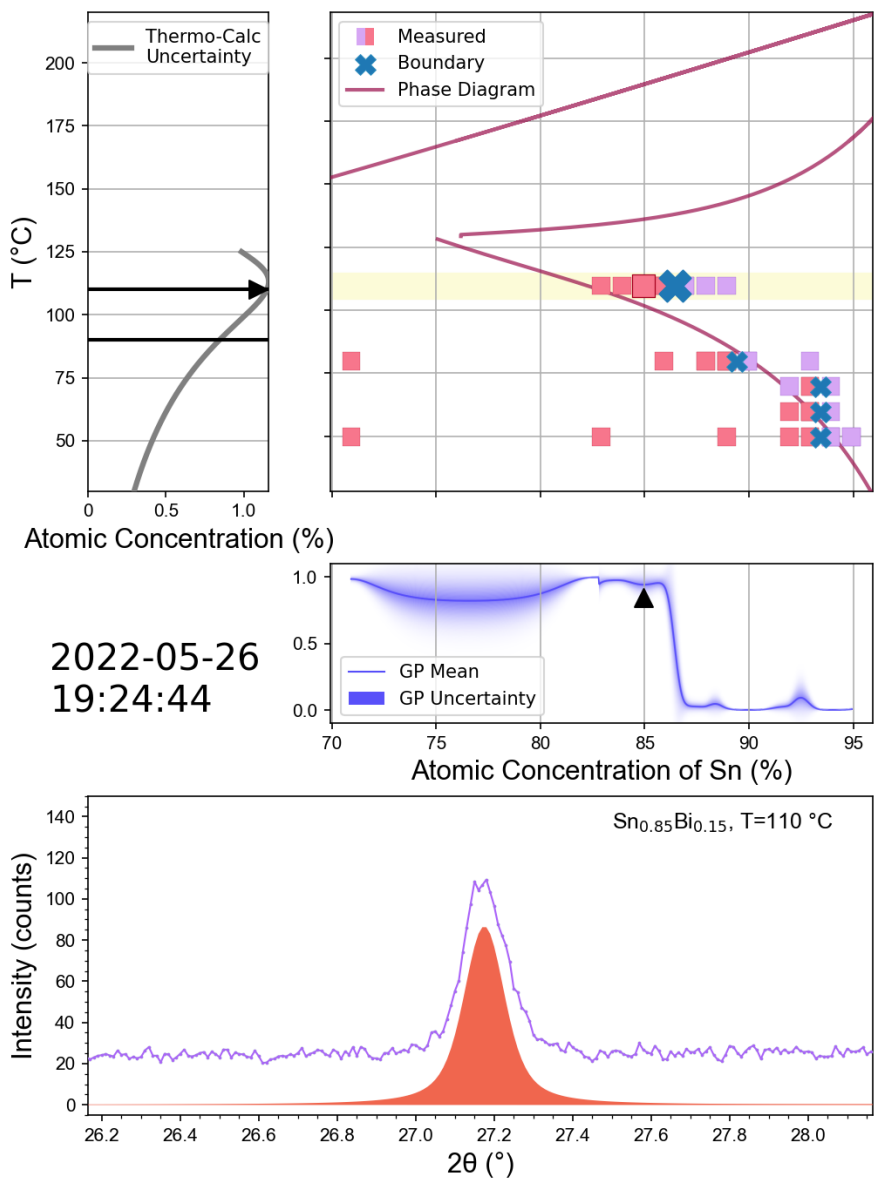

Fig. 4. AMASE at work. (A timelapse video showing the measurement process is included in Supplementary Section 8) The top right panel shows our experimentally measured data points with red color representing the β-Sn + Bi phase and purple color representing the pure Sn phase. The left side panel shows our TC uncertainty values at each temperature, and two horizontal lines show the allowed search range or temperature window. The TC uncertainty

is the standard deviation of the sampled phase boundary composition at each temperature. The middle panel plots the posterior mean and uncertainty prediction of our variational Gaussian process classifier (VGPC). The marker indicates the current measured composition. The XRD pattern and the YOLO + tracker detected peak are shown in the bottom panel, mainly used for visual inspection.

**Fig. 4** shows a snapshot of the time-lapse video capture (video link available in Supplementary Section 8) of one complete AMASE run, which took 8 hours and 22 minutes. The run consisted of 66 XRD measurements carried out at 11 temperature points.

We compare this against various other approaches of rapidly mapping a thin-film phase diagram. An exhaustive "grid measurements" of the spread from 50 °C to 200 °C with 10 °C increment would have required 400 XRD measurements, which would have taken over 60 hours of continuous run. Given the volatile nature of Bi and Sn thin films, such an extended run in air at elevated temperatures would have "damaged" the Sn-Bi film significantly through evaporation, converting the film into off-stoichiometric oxides, etc.

With this time saving, any possible remaining damage/modification of the film, namely extra oxide formation at the surface, is below the detection limit of the diffraction in a 5 min set for each measurement. Under the current experimental condition, we estimate the thickness of such an oxide layer to be less than 5-10 nm. This amount (equivalent volume fraction of 1-2%) is not expected to affect the solved phase diagram constructed for ≈0.5 micron thick films.

We have also considered a variation of the current AMASE implementation, where we perform GP only to navigate the same given composition spread, and then apply CALPHAD afterwards to extrapolate to the entire phase diagram. As discussed in a detailed comparison below, but we see that it is not better than the current interleaving GP and CALPHAD-driven approach.

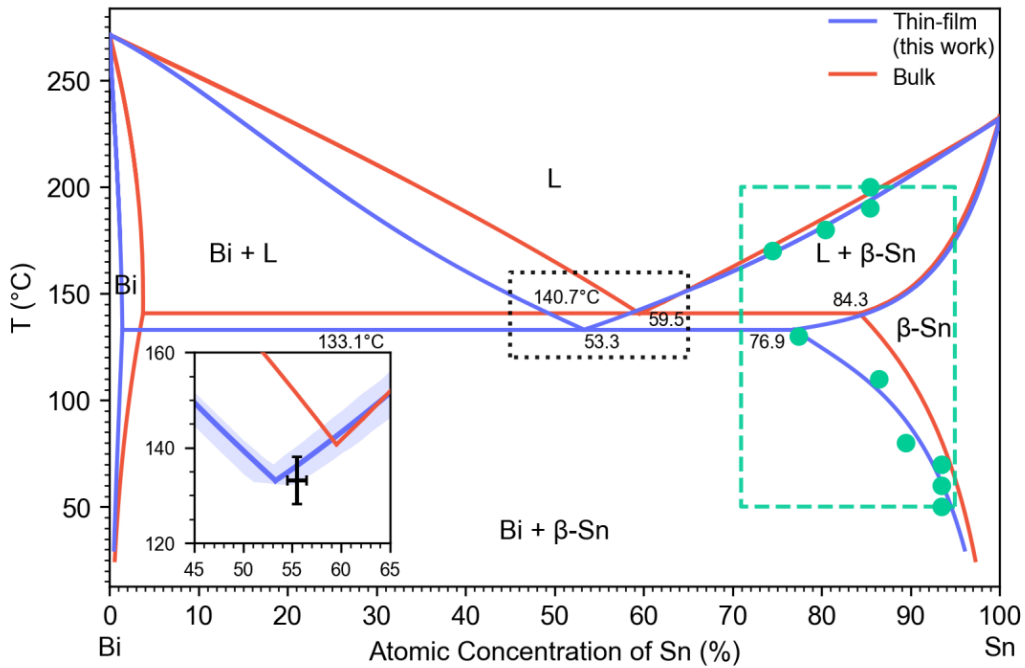

Fig. 5. The comparison of bulk vs. the AMASE thin-film phase diagram. The region inside the green dotted line is the phase space accessed by the AMASE process, with the green dots denoting the phase boundaries experimentally determined by the process. L stands for the liquid phase. The thin-film Sn-Bi phase diagram obtained by the AMASE live run is shown in blue lines, while the bulk Sn-Bi phase diagram is given in red lines(*32*). (Inset) Zoomed-in comparison of bulk vs. thin-film phase diagram near the eutectic point, together with the experimentally determined eutectic point. The boxed region is marked with a black dotted line in the main plot. The blue shadow is the prediction envelope of the AMASE phase diagram extracted from calculated phase diagrams sampled 50 times. Here, the envelope shows the minimum and maximum temperature at each composition of the sampled phase diagrams. The error bar of the experimentally determined eutectic point indicates uncertainty from the WDS composition measurement and the grid size of the grid search experiment. An overlap between the error of the eutectic point measurement and the envelope indicates good agreement between the AMASE prediction and the experiment.

**Fig. 5** shows the AMASE-generated thin-film phase diagram, which indeed exhibits a significant deviation from the known bulk phase diagram(*32*). The region within the green box is the phase space sampled by the current run. The deviation between the bulk and thin-film phase diagrams can be explained by the additional contribution to the Gibbs free energy from surface tension of small grains(*44*), which are widely reported in nanoscale binary systems such as Ag-Cu nanoparticles(*45*), and Al/Me/Al films (Me= In, Sn, Bi, Pb)(*46*). In order to verify the generated thin-film phase diagram, we have performed an independent determination of the eutectic point with a separate thin-film spread sample focused near the $x = 0.5$ composition. We applied the same YOLO + tracker method to identify the Bi and β-Sn phases in the phase mixture region and mapped the liquidus lines in proximity of the eutectic point. As shown in the zoomed-in inset of **Fig. 5**, the experimentally determined eutectic point in this separate thin-film experiment is found at a Sn mole percent of 55.5 % ± 1.5 % ($x = 0.555 ± 0.015$) and a temperature of 133.2 °C ± 10 °C (406.4 K ± 10 K), which is within 3 % of the AMASE-predicted eutectic point of 53.3 % ± 2% ($x = 0.533 ±$ 0.02), 133.1 °C ± 1 °C (406.3 K ± 1 K), in contrast to the bulk eutectic point of 59.5 % (x = 0.595) at 140.7 °C (413.8 K). There is good agreement between the AMASE phase diagram and the validation experiment.

## Discussion

The AMASE setup is applicable to investigating general systems that show deviation in the free energy landscape going from bulk to thin-film systems. The VGPC + CALPHAD prior provides a flexible output that can easily represent mixture and/or pure phases via multiclass binary classification. The Gibbs phase rule is softly constrained in the CALPHAD prior, which allows the system to work not only in ideal samples but also in scenarios such as the formation of a metastable phase. The utility of VGPC as the central framework brings flexibility for modeling complex phase diagrams with line compound(s) or isostructural regions that can break the continuity of the phase field. The foundational assumption of the workflow is the availability of the bulk phase diagram (i.e., it has free energy functional forms of the bulk system). With modification of the acquisition function, the AMASE workflow can be applied to more complex systems, such as ternary phase diagrams and certain regions of multicomponent systems that are of interest to the broad materials science community.

While detecting an unknown/new phase has been shown to be possible in many previous studies(*47–49*), it is generally difficult to characterize the crystal structure from thin-film XRD data. It is also hard to kick-start a new phase in CALPHAD assessment when thermodynamic properties are lacking. In future work, a machine learning-based phases recognition system could be employed alongside the AMASE system to actively monitor phase composition. Such a system could replace the YOLO peak detection + tracking workflow if it is shown robust enough. In addition, the predictions from first-principles calculation databases such as the Materials Projects and Aflowlib can be linked or pre-loaded into the AMASE system to anticipate potential new phases.

There have been reports of several autonomous workflows for similar purposes, i.e., materials optimization, and some of them had featured inner loops(*25*, *50*), but to the best of our knowledge, the AMASE workflow is the only one which integrates computational and experimental tasks with demonstrated operations in live runs. A key feature of our multi-loop workflow is that updating CALPHAD calculation from the main loop (performed every 40 min) is also used to guide the inner Phase Boundary Search Loop.

To further demonstrate the advantage of having an interleaved interaction between the computational physical model (CALPHAD) and the experimental determination of phase boundaries, we have developed a separate workflow where tracking and choosing the phase boundary is entirely performed by a baseline GP approach. We have extracted key statistical metrics to compare the performance of this approach vs AMASE (Figure 6). Each method was simulated 20 times using the final derived phase diagram from the AMASE live run. Here, the GP approach is explicitly programmed to search for solvus and liquidus lines. Methods are benchmarked against the total number of points used to navigate through the phase diagram and the number of measured points per temperature. As seen in Fig. 6(a), the total number of measured points with the AMASE method is significantly lower than that for the GP approach. We attribute this difference to the ability of the CALPHAD to extrapolate the phase diagram features properly. The GP is found to behave more stochastically in comparison to the AMASE method, indicated by its wider distribution of the total measured points and a larger tail in the distribution of measured points per temperature (Fig 6(b) & (c)). The high frequency observed for bin 0 in Fig 6 (b) is due to the fact that AMASE is able to "skip" many temperature points because of the basic extrapolative nature of CALPHAD, which is simply not possible with the GP approach.

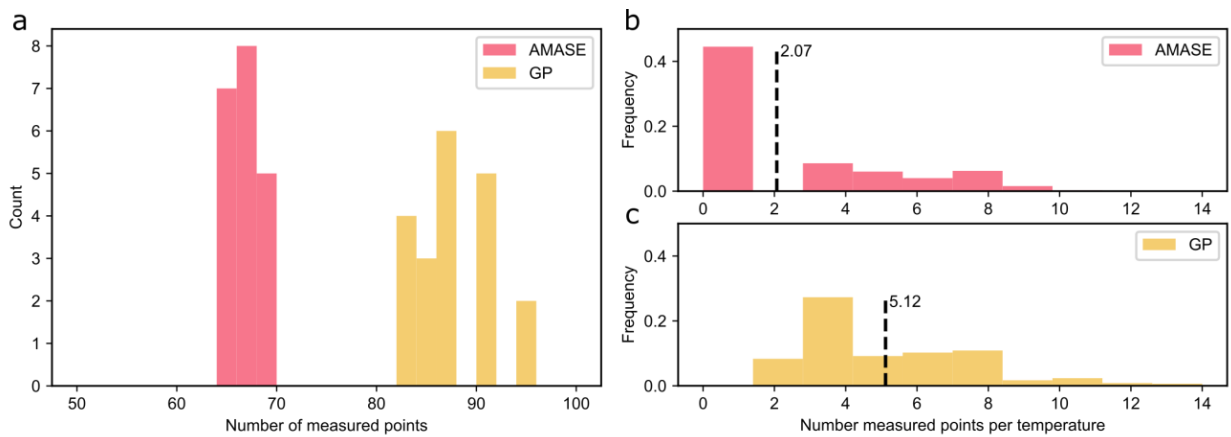

Fig. 6. Performance comparison between AMASE and GP workflow in two statistical parameters. a, The histogram of the total number of measured points for each method took to navigate through the simulated phase diagram for 20 simulated runs. b and c, The histogram of the number of measured points per temperature of the phase-boundary-search-loop for the AMASE and GP workflow, respectively. Dashed lines indicate the average number of measured points of the histogram.

## Conclusion

We have demonstrated an entirely new mode of materials exploration, where experimental and computational tasks are intertwined autonomously in a timed cyclical manner continually informing each other. The phase diagram of the thin-film Bi-Sn binary system is rapidly constructed with the proposed AMASE workflow and the predicted eutectic temperature is verified by a follow-up experiment. As demonstrated here, this allows researchers to probe regions of multi-dimensional parameter space which are outside the reach of physical conditions of typical investigations. Exciting natural extensions of the AMASE operation here include incorporation of other types of computational work such as density functional theory and rapid autonomous mapping of convex hulls, as well as coupling of threaded experiment-theory cycles in research areas beyond materials science. Our work also shows that one can create complete autonomous materials exploration systems solely based on relatively inexpensive existing tools in the lab, thus demonstrating the utility and impact of autonomous science for the general public.

## Data and the code availability

Data that supports the finding of this study as well as the code have been made available as a part of the manuscript.

## Acknowledgements

This project was funded by ONR MURI N00014-17-1-2661, AFOSR FA9550-14-10332, and NIST 70NANB17H301.

## Author contributions

I.T., A.G.K., and J.Z. conceived the project. I.T. supervised the experiments. H.L. wrote the code for the experiment and analysis. D.K. and R.P. fabricated the composition-spread samples. H.L. and H.Y. carried out the autonomous variable temperature x-ray experiment. C.W performed the CALPHAD modeling under the supervision of J.Z. I.T., H.L., C.W., J.Z., A.G.K., and A.M. analyzed the results. I.T., H.L., C.W. and J.Z. wrote the manuscript with input from all authors.

## Disclaimer

The identification of any commercial product or trade name does not imply endorsement or recommendation by the National Institute of Standards and Technology (NIST). These opinions, recommendations, findings, and conclusions do not necessarily reflect the views or policies of NIST or the United States Government.

Perkins, Z. Yang, P. K. Herring, M. Aykol, S. J. Harris, R. D. Braatz, S. Ermon, W. C. Chueh, Closed-loop optimization of fast-charging protocols for batteries with machine learning. *Nature* **578**, 397–402 (2020).

10. S. Sun, A. Tiihonen, F. Oviedo, Z. Liu, J. Thapa, Y. Zhao, N. T. P. Hartono, A. Goyal, T. Heumueller, C. Batali, A. Encinas, J. J. Yoo, R. Li, Z. Ren, I. M. Peters, C. J. Brabec, M. G. Bawendi, V. Stevanovic, J. Fisher, T. Buonassisi, A data fusion approach to optimize compositional stability of halide perovskites. *Matter* **4**, 1305–1322 (2021).

11. M. M. Noack, P. H. Zwart, D. M. Ushizima, M. Fukuto, K. G. Yager, K. C. Elbert, C. B. Murray, A. Stein, G. S. Doerk, E. H. R. Tsai, R. Li, G. Freychet, M. Zhernenkov, H.-Y. N. Holman, S. Lee, L. Chen, E. Rotenberg, T. Weber, Y. Le Goc, M. Boehm, P. Steffens, P. Mutti, J. A. Sethian, Gaussian processes for autonomous data acquisition at large-scale synchrotron and neutron facilities. *Nat. Rev. Phys.* **3**, 685–697 (2021).

12. R. Garnett, *Bayesian Optimization* (Cambridge University Press, 2023).

13. M. Abolhasani, E. Kumacheva, The rise of self-driving labs in chemical and materials sciences. *Nat. Synth.* **2**, 483–492 (2023).

14. M. B. Rooney, B. P. MacLeod, R. Oldford, Z. J. Thompson, K. L. White, J. Tungjunyatham, B. J. Stankiewicz, C. P. Berlinguette, A self-driving laboratory designed to accelerate the discovery of adhesive materials. *Digit. Discov.* **1**, 382–389 (2022).

15. B. P. MacLeod, F. G. L. Parlane, T. D. Morrissey, F. Häse, L. M. Roch, K. E. Dettelbach, R. Moreira, L. P. E. Yunker, M. B. Rooney, J. R. Deeth, V. Lai, G. J. Ng, H. Situ, R. H. Zhang, M. S. Elliott, T. H. Haley, D. J. Dvorak, A. Aspuru-Guzik, J. E. Hein, C. P. Berlinguette, Self-driving laboratory for accelerated discovery of thin-film materials. *Sci. Adv.* **6**, eaaz8867 (2023).

16. L. M. Roch, F. Häse, C. Kreisbeck, T. Tamayo-Mendoza, L. P. E. Yunker, J. E. Hein, A. Aspuru-Guzik, ChemOS: An orchestration software to democratize autonomous discovery. *PLoS One* **15**, e0229862 (2020).

17. B. Burger, P. M. Maffettone, V. V Gusev, C. M. Aitchison, Y. Bai, X. Wang, X. Li, B. M. Alston, B. Li, R. Clowes, N. Rankin, B. Harris, R. S. Sprick, A. I. Cooper, A mobile robotic chemist. *Nature* **583**, 237–241 (2020).

18. R. W. Epps, M. S. Bowen, A. A. Volk, K. Abdel-Latif, S. Han, K. G. Reyes, A. Amassian, M. Abolhasani, Artificial Chemist: An Autonomous Quantum Dot Synthesis Bot. *Adv. Mater.* **32**, 2001626 (2020).

19. S. Langner, F. Häse, J. D. Perea, T. Stubhan, J. Hauch, L. M. Roch, T. Heumueller, A. Aspuru-Guzik, C. J. Brabec, Beyond Ternary OPV: High-Throughput Experimentation and Self-Driving Laboratories Optimize Multicomponent Systems. *Adv. Mater.* **32**, 1907801 (2020).

20. S. Fujino, M. Murakami, V. Anbusathaiah, S.-H. Lim, V. Nagarajan, C. J. Fennie, M. Wuttig, L. Salamanca-Riba, I. Takeuchi, Combinatorial discovery of a lead-free morphotropic phase boundary in a thin-film piezoelectric perovskite. *Appl. Phys. Lett.* **92**, 202904 (2008).

21. J. Cui, Y. S. Chu, O. O. Famodu, Y. Furuya, J. Hattrick-Simpers, R. D. James, A. Ludwig, S. Thienhaus, M. Wuttig, Z. Zhang, I. Takeuchi, Combinatorial search of thermoelastic shape-memory alloys with extremely small hysteresis width. *Nat. Mater.* **5**, 286–290 (2006).

(2020).

# Supplementary Materials

## 1. YOLO-based model

Convolutional neural networks (CNN) have proven to be effective in performing complex object detection. Among the many architectures, the YOLO architecture(*1, 2*) has a simplistic form while still having versatile performance and low computational cost. The idea of YOLO is to attach a detection head that transforms activations of a hidden layer in a CNN to the bounding box coordinates, an objectiveness score, and other auxiliary properties such as classes. Here, we employ a U-Net-like(*3*) CNN model as the backbone of the YOLO model, and the activation of its last hidden layer is used to create the prediction. We choose the box anchor style for our bounding box so that the model predicts the offset of the center and width of all the 1D bounding boxes. When training the model, given a set of ground truth bounding boxes, we assign them to the predicted bounding box by a function B:

$$B(f_i(x), y) = \mathrm{argmax}_{y_j}\left(\mathrm{IOU}(f_i(x), y_j)\right) \quad (1)$$

The Interception over Union (IOU) for 1D box is defined as:

$$\mathrm{IOU}(b_1, b_2) = \max\left(0, \max\left(b_{1c} + \frac{b_{1w}}{2}, b_{2c} + \frac{b_{2w}}{2}\right) - \min\left(b_{1c} - \frac{b_{1w}}{2}, b_{2c} - \frac{b_{2w}}{2}\right)\right) \quad (2)$$

The final loss function is:

$$L(x, y) = \sum_{i \in S}(|B(f_i(x), y)|_1^1) + \sum_{i \in S} g_i(x)\log(g_i(x)) + \sum_{i \notin S}(1 - g_i(x))\left(\log(1 - g_i(x))\right) \quad (3)$$

Where $x$ is the input pattern as a vector, $y$ are the true bounding boxes (a matrix of multiple rows of box center and box width), $f_i(x)$ is the i-th bounding box prediction, and $g_i(x)$ is the confidence score of i-th the predicted bounding box, $y_j$ is the j-th ground truth bounding box, $b_i$ is the i-th bounding box parameterized by its center $b_{ic}$ and width $b_{iw}$. The model was trained on 20000 simulated randomly distributed Voigt peaks for 10 epochs with a linear learning rate schedule ($1 \times 10^{-3}$, $1 \times 10^{-5}$). The parameter range that characterizes three different modes of Voigt peaks is shown in **Table 1**:

Table 1. Voigt peaks parameter ranges for three different modes.

|  | Amplitude (a.u.) | Sigma (°) | Gamma (°) | Height (a.u.) | FWHM (°) | Center (°) | Number of Peaks |
|---|---|---|---|---|---|---|---|
| Crystalline | 0.0066-2.661 | 0.027-0.555 | 0.027-0.555 | 0-1 | 0.1-2 | 20-120 | 0-15 |
| Amorphous | 0.665-3.991 | 1.388-4.165 | 1.388-4.165 | 0-0.3 | 3-10 | 25-40 | 0-5 |
| Substrate | 0.007-1.331 | 0.0003-0.0138 | 0.0003-0.0138 | 5-20 | 0.01-0.05 | 20-120 | 0-5 |

The simulated patterns have a 2θ range of 20 ° to 120 ° with a 0.1 ° step size. Gaussian noise and background profile (a second-order polynomial) are dynamically generated and added to the simulated pattern during training.

The trained model is characterized by the mean Average Precision (mAP) metric, which measures how much the predicted bounding box overlaps with the true box and the reliability of each prediction. The higher the mAP, the higher the performance. The mean Average Precision of the YOLO model is calculated to be 0.78, indicating an excellent performance over the task considering general mAP is around 0.6 for the COCO dataset(*4*). The Non-Maximum Suppress (NMS) algorithm(*5*) is used to select the final bounding box from the raw prediction. The NMS algorithm works by iteratively selecting the remaining most confident boundary box prediction and removing nearby boxes with IOU higher than the defined NMS threshold. The process repeats until all the remaining box has a confidence level (0 to 1, where 0 is low confidence and 1 is high confidence) lower than the defined threshold. Here, we set the IOU threshold level to 0.1 and confidence level to 0.5 during the inference. After applying the NMS, the most certain predictions are drawn from models and are ready to be used as a prior to fit the x-ray diffraction (XRD) pattern. The two parameters of the NMS algorithm can be tuned based on the transferred tasks or the testing part of the training data.

## 2. U-Net Active Noise Cancelling

The presence of noise is inevitable when dealing with experimental diffraction data. Traditional workflows use filtering or interpolation to separate the smooth part from the spiky noise. The signal sharpness and the intensity usually change during the process, which eventually introduces a new set of artifacts. Here, we used a 1D U-Net(*3*) to predict the actual noise pattern given the raw XRD pattern. To implement such a model, a self-supervised scheme in which we let the model predict the undistorted pattern from the noisy input. The loss function is given by:

$$L(x, x') = \left|(x' + f(x',\theta)) - x\right|_1^1 \quad (4)$$

where $x'$ is the noise-augmented diffraction pattern, $f(x',\theta)$ is the generated anti-noise pattern from U-Net with parameters $\theta$, and the $x$ is the smooth simulated ground true pattern. Smooth spectrums are generated from inorganic crystal structural database (ICSD)(*6*) to mimic the actual data distribution of the XRD patterns. Noise patterns are sampled from an independent Gaussian distribution. The model is trained on 69500 simulated patterns for five epochs. The average residual of prediction in the validation dataset is calculated to be $2 \times 10^{-5}$.

## 3. Raw Diffraction Pattern Analysis

3.1 U-Net Noise Extraction

Besides extracting the main features from the diffraction pattern, the underlying noise that follows the Poisson statistics can tell us the quality of the measurement. Instead of recognizing the noise from the post-fitted residual pattern, we used a U-Net model (see Section 2 for details of the model)(*7*) with the same backbone as the YOLO model to predict the noise pattern from raw diffraction patterns. The training dataset is generated from an inorganic crystal structural database (ICSD)(*6*) with Topas software(*8*), and noise is generated from the Gaussian distribution. This model is transferred to a real environment without further training. The global noise level is then extracted from the estimated noise pattern and used as a factor to scale the diffraction pattern with some predefined noise-to-signal ratio in mind. Here, the noise-to-signal ratio is set to be 50 for the best model performance. An example using the same XRD pattern is shown in **Fig. S1a**. The anti-noise pattern generated from our model correctly describes the distribution of the actual noise signal. The smoothed pattern, created by adding the anti-noise pattern to the original pattern, retains all the sharp features from the original pattern, allowing better estimation of the noise level and visualization.

3.2 YOLO Peak Detector

XRD patterns of polycrystalline thin-film materials carry information about the existence of particular phases. Thus, by extracting individual diffraction peaks from the XRD pattern and matching them with peaks calculated from reference structures, one can tell whether peaks appear or not. The extracting step has non-trivial solutions due to peak overlapping, broadening, and shifting. Using the conventional peak finding algorithm based on searching local maximum would fail at heavy peak overlapping region and/or broad peaks from poor crystalline samples. Previously, neural networks have been used to classify symmetry groups, prototype structures, or phases from raw XRD patterns(*9–11*). Data-driven methods of phase mapping have been developed for scenarios with a sufficient number of XRD patterns (e.g., from a combinatorial library), with non-negative matrix factorization (NMF)(*12*) or generative adversarial network (GAN)(*13*). However, none of these methods can provide information for individual peaks when the number of XRD pattern is low. Here, we propose using the 1D YOLO convolutional neural network(*1, 14*) to bridge the gap for machine learning-based phase mapping methods (see Section 1 for details of the model). Given a min-max normalized diffraction pattern, the models output multiple pairs of peak centers and widths that enclose the diffraction peaks where the maximum is the product of the extracted noise level and the user-defined ideal signal-to-noise ratio. The model was trained on simulated patterns with randomly distributed Voigt peaks. The dataset contains various modes that cover the materials from poor/good crystalline phases, substrate, and amorphous peaks. The model was successfully transferred to a real environment achieving transfer learning(*15*). At each measurement, we apply the YOLO network in conjunction with the non-maximum suppression (NMS) algorithm(*5*) to detect peak positions with high prediction confidence. These peak positions would be used as priors for fitting a numerical model, which further refines the peak positions and shape parameters. An example of the YOLO model's output is shown in **Fig. S1b**, where the red rectangle shows the predicted bounding boxes' centers and widths. The predicted bounding boxes cover all observable peaks from human eyes, and widths (2 × FWHM) of peaks are accurately captured.

3.3 Numerical Fitting

For an accurate extraction of peak parameters, one may need to rely on numerical models. The diffraction peaks are modeled by Voigt peaks, and their parameters are initialized by predictions from the YOLO neural network. The background signal is modeled by a Chebyshev Polynomial and is initialized from regions with no peaks. The degree of the Chebyshev Polynomial is randomly sampled from 4 to 7. The numerical model is optimized via the Levenberg–Marquardt algorithm implemented by Scikit-learn and lmfit python libraries(*16, 17*). The fitting quality is monitored by examining the relative standard deviation of fitted parameters. A new iteration of fitting with a different Chebyshev Polynomial degree would be conducted if the previous run fails to converge. Compared with computationally heavy Bayesian inference approaches(*18, 19*), our method starts at a healthier position, resulting in much less computational overhead during the optimization process. An example of the optimized Voigt peaks and background models is shown in **Fig. S1c**. Using the prior from the YOLO model, the optimization process takes less than a second. The optimized Voigt peaks and the polynomial background fit nicely to the original pattern, providing a solid foundation for the subsequent analysis.

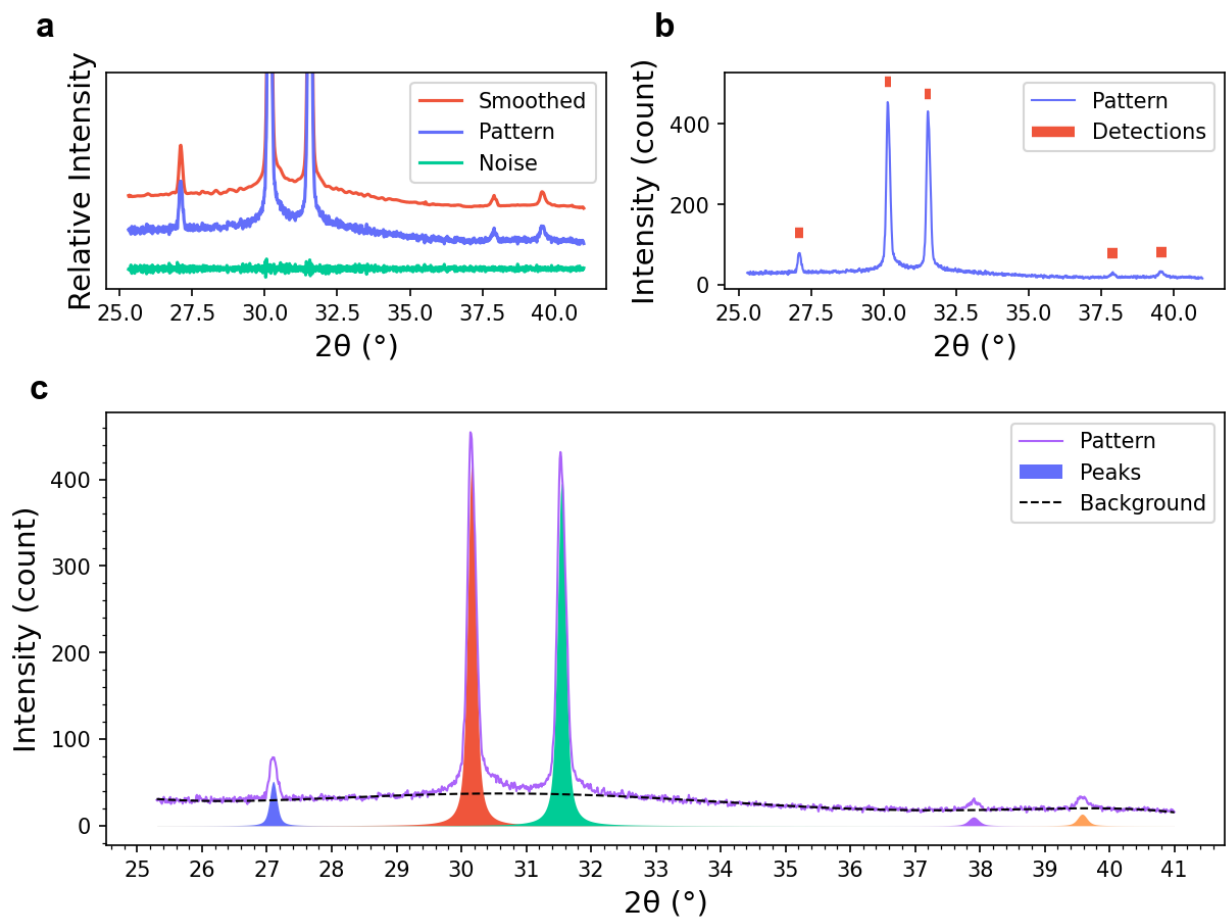

Fig. S1. An example of XRD pattern analysis. A) Output from our U-Net Denoiser. The anti-noise, original pattern and smoothed patterns are shown in different colors. Detected peaks' positions from the YOLO model. The detected peaks' centers and widths are represented by red rectangles' center and width. C) Optimized numerical model of the pattern. The optimized Voigt peaks are shown in shaded regions. The background component is shown in a dashed line.

### 3.4 Phase identification

Indexing the extracted peaks is essential to determine whether a phase is present. For room temperature samples, we can either hand-label or use reference structures from the database to assign the material and plane family to which the peak belongs. To study the peak shift with respect to the position at room temperature as a function of composition and temperature, we employed multiple data-efficient Gaussian process (GP) models to predict the peak positions of previously unknown conditions (composition and temperature). This GP model uses two different kernels on temperature and composition separately. A Matern52 kernel is used to model the complex relationship in the composition space, and a linear kernel is used for the temperature space to model the linear part of the thermal expansion. The final kernel is the product of these two kernels. Here, we use a mean function of a constant value equal to the peak location of the reference structure at room temperature. The target peak is identified if a peak is located in the predicted position and the peak's amplitude is larger than a threshold value. The threshold is usually determined from the noise level extracted before. A peak with an amplitude higher than the noise level is classified as the target phase, and vice versa for the opposite case.

# 4. Thermodynamics-Based Phase Boundary Search

## 4.1 Thermodynamics Models and Uncertainty

By enlisting the automatic analysis of diffraction patterns discussed above, we are able to run autonomous physics-based phase mapping to explore the temperature-dependent phase diagram of a thin-film Bi-Sn binary system. The system is known for its eutectic composition with a low eutectic temperature. The phase diagram exhibits rapid changes around the eutectic temperature, imposing difficulty for a continuous machine-learning model to predict this trend purely from data. This problem is bridged by combining experimental data with theory prediction prior, which comes from CALPHAD (CALculation of PHAse Diagrams) calculations.

CALPHAD is an approximation to the thermodynamic equilibrium theory in the same fashion that the density functional theory (DFT) is an approximation to the quantum mechanics theory. The difference is that the model parameters in DFT have all been assessed beforehand, and no additional fitting is usually required whereas the model parameters in CALPHAD are incomplete and need to be assessed from experimental data system by system. The stability of a material or phase is usually quantified by its Gibbs free energy as a function of temperature, pressure, and composition. A common practice for modeling the Gibbs free energy in binary systems is to use the substitutional solution model. The theoretical thermodynamic model of the Bi-Sn thin-film system is modeled using the Thermo-Calc (TC) software(*20–25*). The substitutional solution model for the molar Gibbs energy of a phase θ in a binary system can be expressed as (*25–27*):

$$G_m^\theta = x_1{}^oG_1 + x_2{}^oG_2 + RT[x_1 ln(x_1) + x_2 ln(x_2)] + x_1 x_2 L_{12} \tag{5}$$

where $x_i$ denotes the mole fraction of element *i*, *R* is the gas constant, ${}^oG_i$ is the Gibbs energy of the pure element *i* whose value can be written as a power series of temperature *T* in the following general equation by selecting a reference state denoted by stable element reference (SER) for element *i* at 298.15K (see Section 6 for detailed expression of SER):

$${}^oG_i = a_0 + a_1 T + a_2 T\ln(T) + a_3 T^2 + a_4 T^{-1} + a_5 T^3 + \cdots, \quad T_1 < T < T_2 \tag{6}$$

$L_{12}$ describes the binary interaction between the two elements, and is expressed as Redlich-Kister polynomial:

$$L_{12} = \sum_{v=0}^{k}(x_1 - x_2)^v \cdot {}^vL_{12} = \sum_{v=0}^{k}(x_1 - x_2)^v \cdot ({}^va_i + {}^vb_i T) \tag{7}$$

Therefore, by plugging Eq. (7) into Eq. (5), the equation of $G_m^\theta$ becomes:

$$G_m^\theta = x_1\,^oG_1 + x_2\,^oG_2 + RT[x_1 ln(x_1) + x_2 ln(x_2)] + x_1 x_2 \sum_{v=0}^{k}(x_1 - x_2)^v \cdot (^va_i + {^v}b_i T) \qquad (8)$$

Due to the microstructure and melting point difference between the thin-film and bulk alloys, the Gibbs energy function of pure elements can vary. Therefore, a single constant parameter is added to the Gibbs energy function of the BCT phase of Sn to account for a change of its melting point, which can better explain the thermodynamic behavior in the thin-film Sn-Bi system. For the binary interaction parameters, only the 0$^{th}$ order ($v = 0$) and 1$^{st}$ order ($v = 1$) in $L_{12}$ are used in the assessment of solid phases, and one more parameter of 2$^{nd}$ order ($v = 2$) is used in assessing the liquid phase; which is quite common for CALPHAD thermodynamic assessments. Since only experimental phase boundary data are available here, we use one adjustable constant $^va_{ij}$ and ignore the $^vb_{ij}$ term in $L_{12}$, which is again quite common during CALPHAD assessments. We thus have two parameters in the solid phase and three parameters in the liquid phase in describing the binary interactions. Three phases are predetermined to exist in the experiment, the body-centered tetragonal (BCT) phase of pure Sn, the rhombohedral (RHO) phase of pure Bi, and the liquid phase. Therefore, 8 adjustable parameters as represented by G(LIQUID, Sn, Bi; 0), G(LIQUID, Sn, Bi; 1), G(LIQUID, Sn, Bi; 2), °G(BCT, Sn), G(BCT, Sn, Bi; 0), G(BCT, Sn, Bi; 1), G(RHO, Sn ,Bi; 0), and G(RHO, Sn, Bi; 1) are employed in the thermodynamic model assessment.

Given experimentally determined phase boundaries (e.g., β-Sn-Bi solvus and β-Sn liquidus line) and Gibbs free energy functions of pure Sn and Bi, the substitutional solution model can be updated through optimization. The updated model is used to evaluate a new Sn-Bi phase diagram which extrapolates phase boundary information into "uncharted" composition and temperature regions. The uncertainty of the new prediction is quantified via Monte Carlo method which is great mean for evaluating uncertainty from an intractable probability distribution. The parameters of the CALPHAD model are converted into i.i.d random variables with a Gaussian distribution. The mean and standard deviation of each random variable is obtained from the VALUE and REL.STD.DEV field from the output of the Thermocalc PARROT optimization operation. We samples multiple set of parameters from the Gaussian distribution and use them to indirectly sample phase diagrams. The uncertainty of the phase boundary could be quantified as:

$$U^\xi(T) = \sum_{i=1}^{n}\left(\tau_i^\xi(T) - \overline{\tau^\xi}(T)\right)^2 \qquad (9)$$

Where $U^\xi(T)$ is the variance-measured of the phase boundary ξ at temperature T, $\tau_i^\xi(T)$ is the *i*-th sampled TC-predicted composition of the phase boundary ξ at temperature $T$ and $\overline{\tau^\xi}(T)$ is the mean of these predictions. The uncertainty describes the discrepancy among n = 10 sampled phase diagrams. To decrease the uncertainty of predictions in the thermodynamic modeling, it is most rewarding to jump to the temperature with the highest uncertainty to perform the next cycle of measurement.

4.2 Active Learning for Phase Boundary Search

By leveraging phase diagram computation using CALPHAD, we deploy a Gaussian Process classifier conditioned on the output of the thermodynamic model to decide the subsequent composition and the system temperature to measure. Using the most recently evaluated phase boundary lines as prior information and experimentally determined phase compositions, we fit a Variational Gaussian Process (VGP) classifier(*28–30*) using the GPFlow Library(*31*) for each phase to predict the most probable phase boundary composition. We construct a custom kernel:

$$K(x_c, x_T, x_p, x'_c, x'_T, x'_p) = \text{Matern32}(x_c, x'_c) \cdot \text{Matern32}(x_T, x'_T) + \text{RBF}(x_p, x'_p) \quad (10)$$

where $x_c$, $x_T$, $x_p$ and $x'_c$, $x'_T$, $x'_p$ are the chemical composition, temperature, and phase information modelled via TC of two data points, respectively. Here, we simplified the formulation of $x_p$ for this binary system into a binary value. For the solvus line, we use $x_p = 0$ to represent the Bi + Sn mixture phase region, and $x_p = 1$ to represent the pure Sn region. Similarly, for the liquidus line, we set $x_p = 0$ for liquid + Sn or pure Sn regions and $x_p = 1$ for the liquid only region. Since it is a binary system, one VGP classifier model is sufficient to describe the phase around the phase boundary of the solvus or liquidus line. One can easily generalize the formulation of $x_p$ into multiclass one-hot encoding where multiple binary variables can be used to represent the state of different phases in the system. Matern32 is the Matérn 3/2 kernel (ν = 3/2) and RBF is the radial basis function kernel(*32*). The VGP classifier model is fitted against the available experimental data and the latest version of the TC assessed phase diagram to update its posterior phase diagram prediction, which used the same formulation as the TC prior.

Using the prediction of the VGP model, samples at the predicted phase boundary position were measured, and the model is iteratively updated until the model prediction converges. At each iteration VGP prediction, a custom acquisition function $\alpha(x) = -|m(x) - 0.5|$ was used to actively search for phase boundary. $m(x)$ is the mean VGP classifier prediction at data point $x$ with composition $x_c$, temperature $x_T$, and phase information $x_p$. The next datapoint to sample is: $x^* = \arg\max_{x \in X} \alpha(x)$, where $X$ is a set of datapoints with temperature $T$. The prediction is converged if the next datapoint $x^*$ didn't get changed in the consecutive iteration. This converged datapoints is considered the phase boundary at the current temperature if its neighboring compositions show drastic changes in phase composition. The phase composition, in this case, is actively monitored by multiple GP trackers discussed in Section 3. If the experimental evidence disagrees with the model prediction, a sample with a randomly selected composition is measured to steer the VGP model out of the convergence. With up-to-the-minute measured phase boundary data, the thermodynamic model is re-adjusted by optimization to match the experimental observation.

Once the phase boundary at the current temperature is experimentally determined, the system turns to measure the phase boundaries at other temperatures. Since raising the temperature would cause irreversible change to the sample, the system temperature is only increased. Here, we select the temperature with the highest $U^\xi(T)$ within a user-defined range as our new system temperature (up to 30 °C above the temperature where the boundary search took place). The subsequent phase boundary composition measured at this new temperature would efficiently reduce the uncertainty in the TC predictions using the reassessed thermodynamic model.

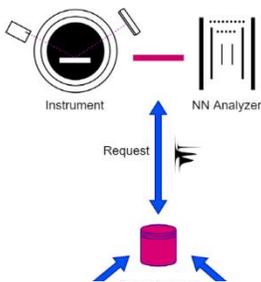

Fig. S2. Detailed workflow of the AMASE software. The process is orchestrated by the main node in the center that handles the storage and data acquisition. The active learning (AL) agent uses the structured data from the main node to guide the experiment while the experiment node collects experimental and Thermo-Calc node provides theoretical prediction.

The schematic of the workflow is shown in **Fig. S2**. The AMASE software is driven by the active learning agent, which acquires input from the database and queries the experiment node for the following most informative data point. The data manager not only stores data points but also coordinates with the XRD instrument and TC to finish the request from the active learning agent. The XRD machine is controlled via a custom-built socket connection, and the raw experimental data is transferred back to the main node for analysis. The aforementioned YOLO-based model, in conjunction with the numerical model, is used to de-convolute the diffraction pattern and extract peak parameters. These parameters, combined with reference structures, provide experimental evidence for our active learning agent and the CALPHAD model to update themselves. At each temperature, the phase boundary is rapidly discovered through XRD by an active learning agent and experimentally confirmed by checking the nearby compositions. The process is terminated when the active learning agent determines there is no phase boundary in the current temperature level. In the current setup, the goal is to discover the liquidus and solvus boundaries on the Sn side of the phase diagram. When scanning through a boundary, instead of increasing the temperature in a fixed step, we sample the predicted phase diagram to locate the temperature with the most uncertain boundary composition and then gradually heat the system toward that temperature at a pace of 10°C/min. When switching from the solvus to liquidus line search, we use the phase diagram predicted by TC (*20*) to search for the temperature at which the liquidus line first appears and directly heat to that temperature. These two approaches further reduce the overall experimentation time with little cost in information lost. Compared with the traditional grid search method, our approach results in ≈6 times speed up and is critical to prevent oxidation in the high-temperature range in air.

## 5. End-point detection scheme

The cyclical process of phase boundary search and temperature adjustment continues until both VGP classifier and the thermodynamic model predicted phase boundaries disappear at higher temperatures. A VGP classifier-guided scanning method was used to experimentally confirm the vanishing of the phase boundary. This prediction works similarly to the VGP classifier model but without the use of the thermodynamic model TC prior. The kernel used in this VGP classifier is:

$$K(x_c, x_T, x'_c, x'_T,) = \text{Matern32}(x_c, x'_c) \cdot \text{Matern32}(x_T, x'_T) \tag{11}$$

where $x_c$, $x_T$ and $x'_c$, $x'_T$ are the chemical compositions and the system temperatures of two data points, respectively. At each iteration, the VGP classifier predicts a plausible phase boundary location, and an XRD measurement is performed to investigate the composition location. If the scanning method finds evidence of a phase boundary, the AMASE will roll back to the search routine and continue the mapping. Otherwise, once VGP classifier predicts there is no phase boundary at the current temperature, AMASE can then switch to a new phase boundary (liquidus) that only appears in higher temperatures, and a new cycle is initiated for this boundary. Here, the modified exploration acquisition function is used:

$$\alpha(x) = \begin{cases} v(x) & \text{if} \quad m(x) - 2\sqrt{v(x)} < 0.5 < m(x) + 2\sqrt{v(x)} \\ 0 & \text{else} \end{cases}$$

where $m(x)$, $v(x)$ is the mean and variance of VGP classifier posterior prediction at data point $x$ with composition $x_c$, temperature $x_T$, and phase information $x_p$. This acquisition is designed guide the agent to explore all the possible phase transitions region before signaling a termination.

# 6. Database of Gibbs energy functions of pure Sn and Bi

The database of Gibbs energy function of pure Sn and Bi is adapted from Scientific Group Thermodata Europe (SGTE) data(*33*). Here is each of the function we used:

(298.15 K < $T$ < 505.07 K)
$$\text{GHSERSN} = -5855.135 + 65.443315\,T - 15.961\,T\ln(T) - 0.0188702\,T^2 + 3.121167\,10^{-6}\,T^3 - 61960\,T^{-1}$$
$${}^oG_{Sn}^{bct} - H_{Sn}^{SER} = \text{GHSERSN}$$
$${}^oG_{Sn}^{rho} - H_{Sn}^{SER} = \text{GHSERSN} + 2035$$
$${}^oG_{Sn}^{liquid} - H_{Sn}^{SER} = \text{GHSERSN} + 7103.092 - 14.087767\,T + 1.47031\,10^{-18}\,T^7$$

(505.07 K < $T$ < 800 K)
$$\text{GHSERSN} = +2524.724 + 4.005269\,T - 8.2590486\,T\ln(T) - 0.016814429\,T^2 + 2.623131\,10^{-6}\,T^3 - 1081244\,T^{-1} - 1.2307\,10^{25}\,T^{-9}$$
$${}^oG_{Sn}^{bct} - H_{Sn}^{SER} = \text{GHSERSN}$$
$${}^oG_{Sn}^{rho} - H_{Sn}^{SER} = \text{GHSERSN} + 2035$$
$${}^oG_{Sn}^{liquid} - H_{Sn}^{SER} = \text{GHSERSN} + 6971.586 - 13.814383\,T + 1.2307\,10^{25}\,T^{-9}$$

(298.15 K < $T$ < 544.55 K)
$$\text{GHSERBI} = -7817.776 + 128.418925\,T - 28.4096529\,T\ln(T) + 0.012338888\,T^2 - 8.381598\,10^{-6}\,T^3$$
$${}^oG_{Bi}^{rho} - H_{Bi}^{SER} = \text{GHSERBI}$$
$${}^oG_{Bi}^{bct} - H_{Bi}^{SER} = \text{GHSERBI} + 4184.07$$
$${}^oG_{Bi}^{liquid} - H_{Bi}^{SER} = \text{GHSERBI} + 11246.066 - 20.63651\,T - 5.955\,10^{-19}\,T^7$$

(544.55 K < $T$ < 800 K)
$$\text{GHSERBI} = 30208.022 - 393.650351\,T + 51.8556592\,T\ln(T) - 0.075311163\,T^2 + 1.3499885\,10^{-5}\,T^3 - 3616168\,T^{-1} + 1.661\,10^{25}\,T^{-9}$$
$${}^oG_{Bi}^{rho} - H_{Bi}^{SER} = \text{GHSERBI}$$
$${}^oG_{Bi}^{bct} - H_{Bi}^{SER} = \text{GHSERBI} + 4184.07$$
$${}^oG_{Bi}^{liquid} - H_{Bi}^{SER} = \text{GHSERBI} + 11336.26 - 20.810418\,T - 1.661\,10^{25}\,T^{-9}$$

bct : body-centered tetragonal, rho : rhombohedral

# 7. Experimental Setup

The Sn-Bi thin-film combinational spread (0.5 μm thick) was fabricated on a 2 in. (50.8 mm) silicon wafer with a thermal oxide layer (6 μm) by co-sputtering Sn and Bi targets at room temperature for 24 minutes. The Bi and Sn gun power are 12 W and 33 W, respectively. The base pressure is 1.9 x$10^{-10}$ Bar (1.4x$10^{-7}$ Torr). The spread compositions were determined using wavelength dispersive spectroscopy (WDS) with error of ±1 at. %. The combinatorial spread was cut into smaller pieces with the desired composition range and sizes. Bruker C2/D8 Discover powder diffractometer with a high-temperature stage was used to collect diffraction images. The exposure time for each XRD measurement was set to be 5 minutes. The diffraction image was integrated into the 1D diffraction data with the 2θ range from 25.3 ° to 41.0 ° and a step size of 0.01 °. The high-temperature stage operated in air can raise the sample temperature up to 200 °C. The temperature ramp rate is set to be 10 °C/min. 5 minutes hold time is used to wait for the sample to equilibrate before taking any XRD measurement.

## 8. AMASE Live run

8.1 Live run time-lapse video

The video for the AMASE live run is hosted by the BOX platform and is available for download through this URL (https://umd.box.com/s/qtrdtrjsgffwadh6nxchfmjwyu94acz9).

8.2 Live run phase diagram evolution

Snapshots of phase diagram evolution are shown in **Fig. S3**.

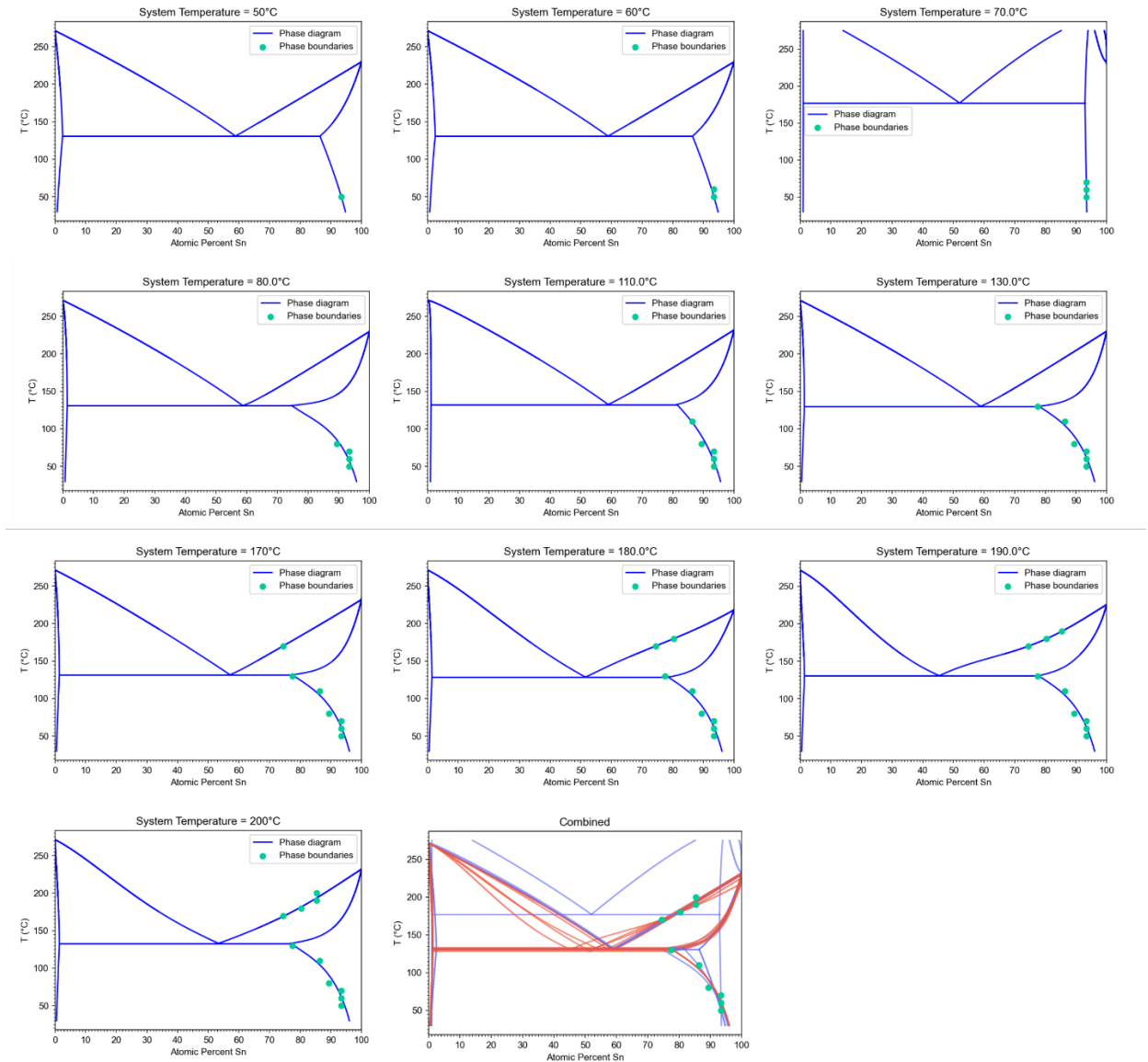

Fig. S3. The evolution of the phase diagram of the thin-film Bi-Sn binary system. First 10 subfigures show the Thermo-Calc assessed phase diagrams at the end of each AMASE temperature scan. The last graph shows all phase diagrams overlaid on top of the others. Blue-to-red color coding is used to show gradual change of temperature from 50 °C to 200 °C (blue to red), respectively.